\documentclass[aip,apl,superscriptaddress,preprint,showpacs]{revtex4-1}

\usepackage{graphicx,textcomp,amssymb,amsmath,dcolumn}
\begin{document}

\title{Photoelectron generation and capture in the resonance fluorescence of a quantum dot}

\author{A.~Kurzmann}
\email{annika.kurzmann@uni-due.de}
\affiliation{Fakult\"at f\"ur Physik and CENIDE, Universit\"at Duisburg-Essen, Lotharstra{\ss}e 1, Duisburg 47048, Germany}
\author{A.~Ludwig}
\author{A. D.~Wieck}
\affiliation{Chair of Applied Solid State Physics, Ruhr-Universit\"at Bochum, Universit\"atsstr. 150, 44780 Bochum, Germany}

\author{A.~Lorke}
\author{M.~Geller}

\affiliation{Fakult\"at f\"ur Physik and CENIDE, Universit\"at Duisburg-Essen, Lotharstra{\ss}e 1, Duisburg 47048, Germany}

\date{\today}

\begin{abstract}

Time-resolved resonance fluorescence (RF) on a single self-assembled quantum dot (QD) is used to analyze the generation and capture of photoinduced free charge carriers. We directly observe the capture of electrons into the QD in an intensity reduction of the exciton transition. The exciton transition is quenched until the captured electron tunnels out of the dot again in the order of milliseconds. Our results demonstrate that even under resonant excitation, excited, free electrons are generated, which can negatively influence the optical properties of a QD. This detrimental effect has been neglected before for dots that are optimized for maximum efficiency and minimum spectral diffusion. 
\end{abstract}

\maketitle

Self-assembled quantum dots (QDs) as artificial atoms in a solid-state matrix show almost ideal properties for optical quantum devices. They are perfect building blocks for single photon sources \cite{yuan2002,kiraz2003quantum, PhysRevLett.108.093602} with a high quantum efficiency \cite{michler2000quantum} and high photon indistinguishability.\cite{santori2002indistinguishable,matthiesen2013,wei2014deterministic} These optical properties arise from very stable exciton transitions.\cite{kuhn2002deterministic,kuhlmann2015transform} However, the ultimate goal of a transform-limited photon stream has not been reached yet, as always spectral wandering of the center frequency of the QD transition is observed. The major source of this spectral jitter is given by charge and nuclear spin noise.\cite{kuhlmann2013,houel2012probing} Interestingly, the generation of free, excited charge carriers and its influence on the optical properties of a single QD has so far been neglected \cite{PhysRevB.72.195339} in resonant optical measurements, like resonance fluorescence (RF). \cite{PhysRevLett.99.187402}

We show here in a time-resolved RF measurement \cite{PhysRevB.81.035332, prl} that free charge carriers are generated in a resonant or close to resonant measurement and that these carriers can be captured by the QD. The capture process is observed by a quenching of the exciton transition. This effect has not been analyzed before, as the probability for a capture process of such an electron into the dot is orders of magnitude lower than the average tunneling rate in previously used QD devices.\cite{PhysRevLett.93.217401, Nick2009} That means, the captured electron is emitted swiftly by tunneling to the charge reservoir and, hence, its influence on the exciton transition is difficult to discern. We show here, that there is always a photoinduced generation of charge carriers and we are able to observe this process in a device that has tunneling times in the order of milliseconds; more than 6 orders of magnitude slower than the exciton recombination time. These photoelectrons are excited from the highly-doped back contact, can relax to the QD layer and influence its optical properties negatively. 

\begin{figure}
  \includegraphics{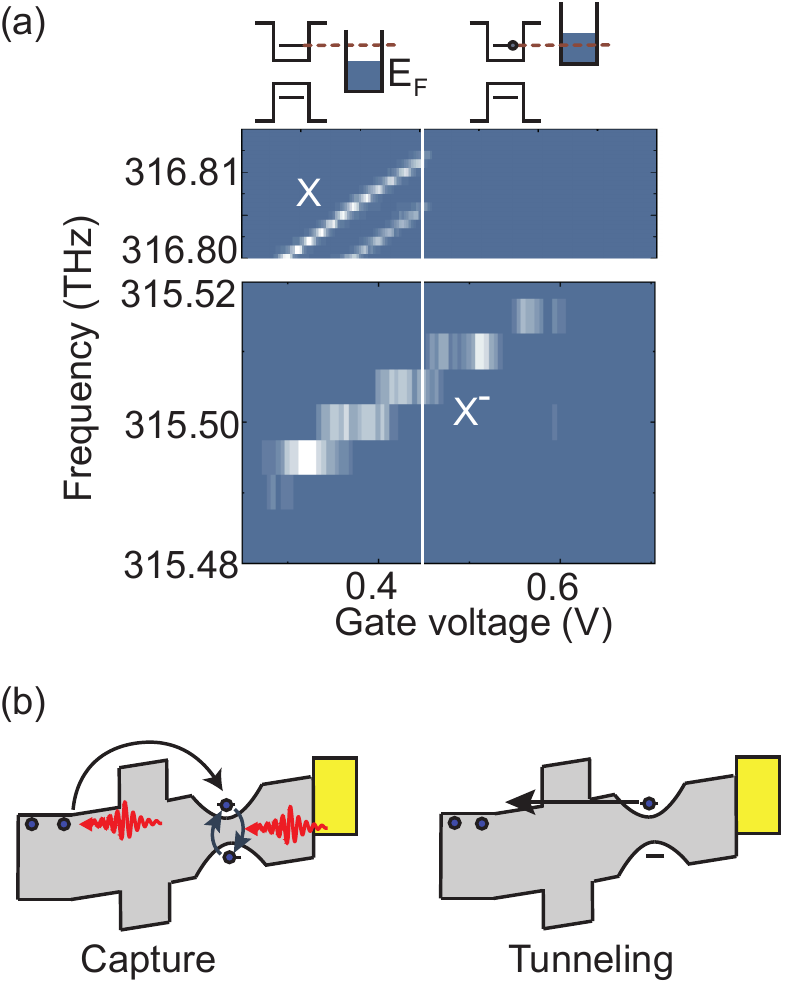}
  \caption{(a) RF of the exciton and trion transition for different laser excitation frequencies and gate voltages. The trion transition is observed for gate voltages, where tunneling into the QD is energetically forbidden. (b) Schematic illustration of the involved processes of generation, capture and tunneling of the optically excited free electrons.}
  \label{fig1}
  \end{figure}
  
   \begin{figure}
      \includegraphics{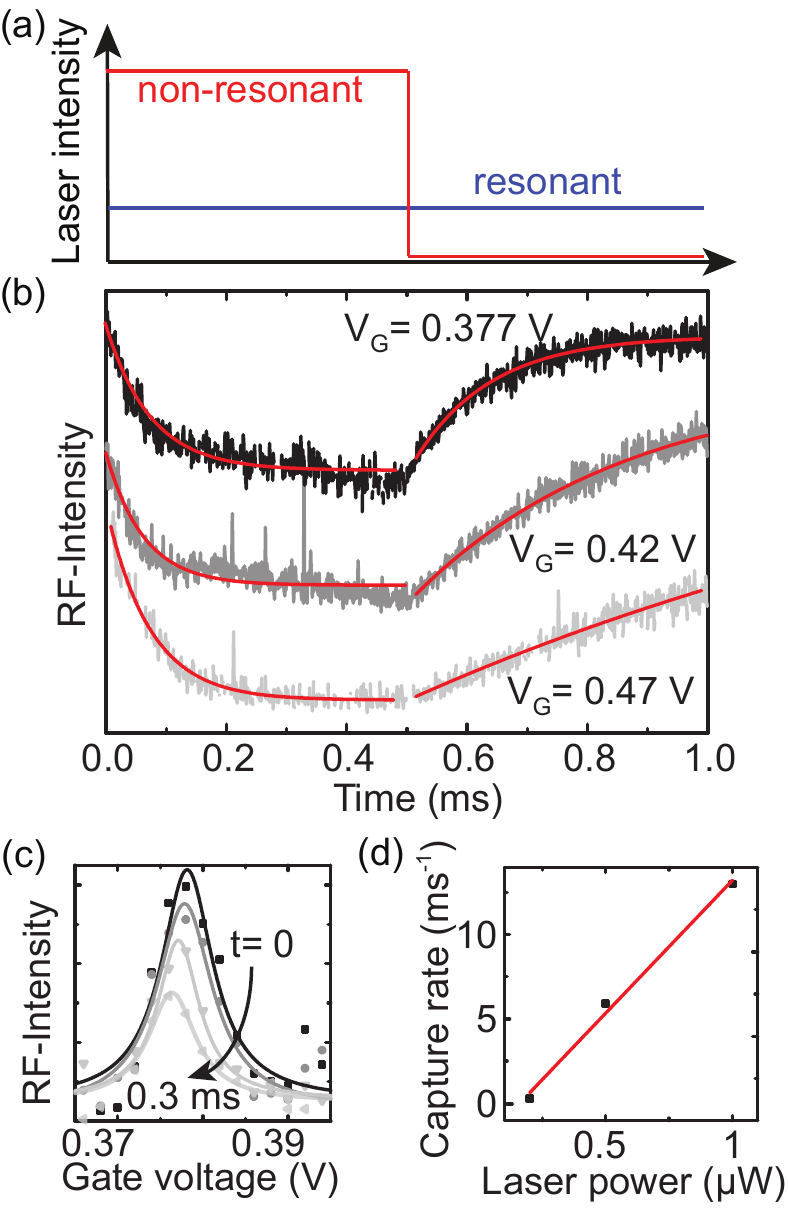}
      \caption{(a) Laser intensity of the two different lasers for the time-resolved RF measurement. The exciton transition is excited resonantly with the same intensity during the measurement. A second laser with a frequency close to the resonance is pulsed with high intensity. (b) Time-resolved RF of the exciton transition for different gate voltages. The RF-signal is quenched, when the non-resonant, second laser is turned on and increases again after switching off the second laser.(c) Evolution of the Lorentzian line shape of the exciton transition after a second non-resonant laser is turned on for a duration of $t=0$, $0.03$, $0.06$, $0.15$ and $0.3\,\text{ms}$. A suppressing of the intensity and an energy shift is observed. (d) Capture rates of electrons into the QD for different laser intensities.}    
      \label{fig2}
      \end{figure}

The investigated sample was grown by molecular beam epitaxy (MBE) and resembles a Schottky diode with a layer of self-assembled InAs QDs. In detail, a $300\,\text{nm}$ GaAs layer was deposited on a semi-insulating GaAs substrate, followed by a $50\,\text{nm}$ silicon-doped GaAs layer, which forms the electron reservoir. A tunneling barrier, consisting of $30\,\text{nm}$ GaAs, $10\,\text{nm}$ $\text{Al}_{0.34}\text{Ga}_{0.66}\text{As}$ and $5\,\text{nm}$ GaAs, was grown. This layer separates the InAs QDs from the doped back contact. The InAs QDs were formed by growing 1.6 mono-layers of InAs partially capped by $2.7\,\text{nm}$ GaAs and flushed at $600^\circ\text{C}$ for 1 minute to shift the emission wavelength to $\approx 950\,\text{nm}$. They were further capped by a $27.5\,\text{nm}$ GaAs layer, a $140\,\text{nm}$ superlattice (35 periods of $3\,\text{nm}$ AlAs and $1\,\text{nm}$ GaAs) and $10\,\text{nm}$ GaAs.  The Ohmic back contact was formed by AuGe and Ni evaporation and annealing. Transparent Schottky gates are prepared on the sample surface by standard optical lithography and deposition of $7\,\text{nm}$ NiCr. On top of these gates, a zirconium solid immersion lens (SIL) is mounted to improve the collection efficiency of the QD emission.\cite{Gerardot2007} A gate voltage applied between the top gate and the Ohmic back contact induces an external electric field and controls the charge state in the QD.\cite{Petroff2001, drexler1994}

We use a confocal microscope setup in a bath-cryostat at a temperature of $4.2\,\text{K}$. For the RF measurements, the exciton ($X$) or trion ($X^-$) transitions are driven resonantly by a linearly polarized and frequency stabilized tunable diode laser. In a confocal geometry, both laser excitation and QD emission are guided along the same path, using a 10:90 beam-splitter. Single QD resolution is archived by a 0.65 NA objective lens in front of the above mentioned SIL, resulting in a spot size of $1\,\mu\text{m}$. The emission of the QD is collected behind a polarizer, which is polarized orthogonally to the excitation laser and suppresses the laser light by a factor of $10^7$. The RF signal of the QD is detected by an avalanche photo diode (APD) and is recorded using a time-to-digital converter with a time resolution of $81\,\text{ps}$.

We observe the exciton transition (X) with its fine structure splitting \cite{gammon1996fine} and quantum confined stark effect\cite{Li2000} in the RF scan in Fig.~\ref{fig1}(a)  The exciton is observed for gate voltages below $0.45\,\text{V}$. At this gate voltage the Fermi energy $E_F$ of the back contact is in resonance with the first QD level and tunneling of electrons into the QD is energetically allowed. For higher gate voltages, the QD will be singly charged and the trion transition is observed for a laser frequency of about $315.51\,\text{THz}$. Surprisingly, for high laser excitation powers, we also observe the trion transition below $0.45\,\text{V}$, even though the tunneling of electrons from the back contact into the QD is energetically forbidden.\cite{PhysRevB.72.195339} We can explain this result by the capture process of optically generated free, excited electrons from the back contact into the dot states, as schematically illustrated in Fig.~\ref{fig1}(b). These electrons are generated in by the laser that is used for the resonant excitation and can diffuse to the QD layer and charge the QD even at gate votages where the tunneling is energetically forbidden. A QD charged with one electron cannot be excited on the exciton transition until the electron tunnels back to the charge reservoir, see right hand side of Fig.~\ref{fig1}(b). Note in addition, that we observe the trion transition with a much smaller intensity than the exciton transition, explained by a strong influence of the Auger recombination in QD structures with thick tunneling barriers.\cite{auger}

   \begin{figure}
           \includegraphics[scale=1]{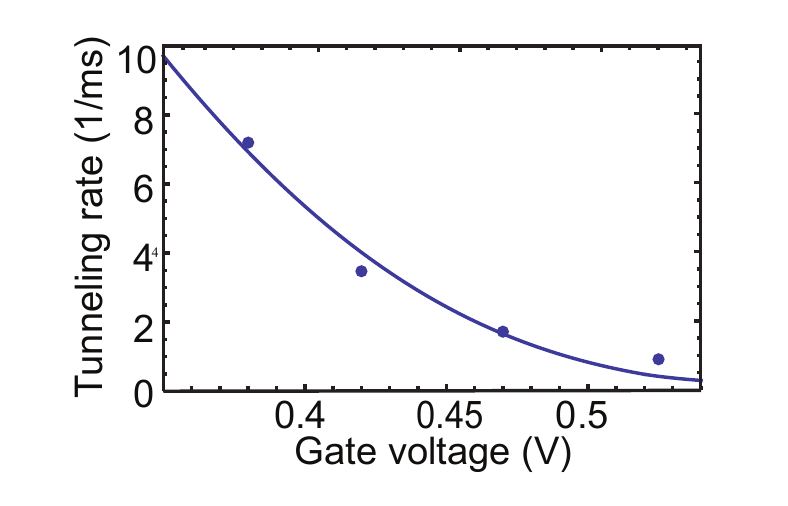}
           \caption{Measured tunneling rates for different gate voltages together with a fit to the data using the WKB approximation for the transparency of the tunneling barrier.}
           \label{fig3}
           \end{figure}

We can directly observe the capture of electrons into the QD in a time-resolved two color n-shot measurement, demonstrated in the following. The first laser drives the exciton transition resonantly at a frequency of $317.803\,\text{THz}$ with a constant low laser intensity (blue line in Fig.~\ref{fig2}(a)), so that no free charges are captured into the dot. To demonstrate that the generation of free electrons is a non-resonant effect, we use a second laser with a frequency of $319.67\,\text{THz}$. This rules out any band-to-band transition by the non-resonant laser excitation. The second laser is used with a pulsed laser intensity that is two orders of magnitude higher than the intensity of the resonant laser. It is switched on at $t=0$, see Fig.~\ref{fig2}(a) (red line) and off at $t=0.5\,\text{ms}$. We repeat this measurement for different detuning between the first laser and exciton energy, so that the Lorentzian line shape of the exciton transition is observed for different times in Fig~\ref{fig2}(c). An intensity reduction and a shift of the maximum of the exciton transition is measured up to a time delay of $0.3\,\text{ms}$. The time evolution of the maximum intensity is shown in Fig.~\ref{fig2}(b). We observe an exponential decay after the laser is switched on at $t=0$ with a relaxation rate of $13\,\text{ms}^{-1}$ (determined from an exponential fit to the data, see red line). The steady state intensity at $t= 0.5\,\text{ms}$ is about 30 percent of the intensity at $t=0$ and it is given by the interplay between electron tunneling and capture. The capture rate can be easily tuned by the laser excitation power, changing the overall relaxation rate into equilibrium; shown in Fig.~\ref{fig2}(d) together with a linear fit to the data (solid red line). We find a linearly increasing capture rate with increasing laser power, as expected for optically excited electrons in the vicinity of the dot. 

In addition to a quenching of the exciton transition, we observe in Fig.~\ref{fig2}(c) a shift of the Lorentzian line to lower gate voltages with increasing time duration. The origin of this shift is not understood yet and needs further investigation. However, it could be due to a accumulation of the photoinduced electrons, changing the band structure and therefore the effective electric field in the QD layer.  

The tunneling rate of electrons out of the dot can be observed for $t>0.5$ ms in Fig.~\ref{fig2}(b). After the non-resonant laser is turned off and the photogeneration of electrons ceases, an  exponential increase of the exciton RF intensity is observed. For increasing gate voltage from $V_G= 0.377 V$ up to 0.47 V we measure a decreasing tunneling rate in Fig.~\ref{fig2}(b) from $7.8\,\text{ms}^{-1}$ down to $1.2\,\text{ms}^{-1}$. We have determined these rates by fitting the tunneling transients exponentially (solid red lines in Fig.~\ref{fig2}(b)) and plotted the values versus gate voltage in Fig.~\ref{fig3}. The tunneling rates decrease with increasing gate voltage as the tunneling barrier is increased and its transparency is reduced (Fowler-Nordheim tunneling). We calculated the transparency of the barrier using the WKB approximation \cite{luyken1999dynamics} and fitted the result to the data points in with an attempt frequency of $f=349\,\text{ns}^{-1}$ as free parameter, shown as solid line in Fig.~\ref{fig3}. The calculation is in very good agreement with the data points. This demonstrates two things: (i) The observed transients in Fig.~\ref{fig2}(b) for $t>0.5$ ms is caused by electron tunneling out of the QD. (ii) More importantly, the quenching of the RF signal for $t<0.5$ ms is a consequence of electron capture into the dot states, that shift the resonance frequency from the exciton to the trion transition. In sub-bandgap excitation, this is possible by photoinduced excitation of free electrons from the highly-doped back contact and subsequent capture in the dot states. 

       \begin{figure}
           \includegraphics[scale=1]{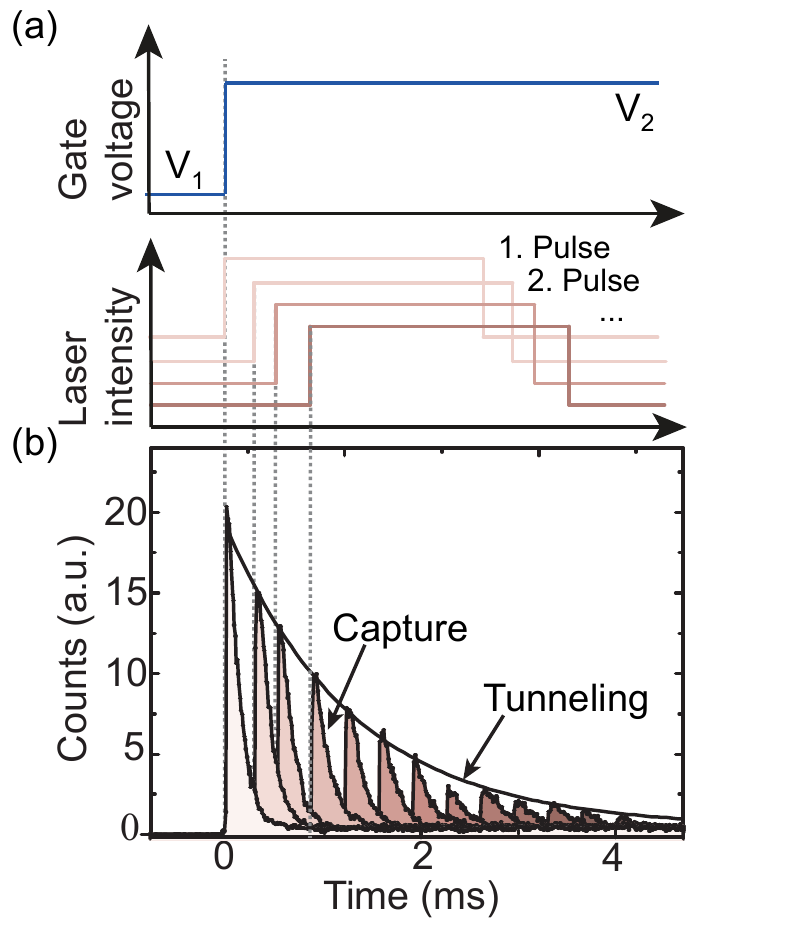}
           \caption{(a) Upper panel: Pulsed gate voltage, with the lower voltage, where the QD is uncharged and the higher voltage, where the QD gets charged. Lower panel: Pulsed Laser intensity with the different time delays used for the measurement in (b). (b) Measurement of the time dependent RF-signal revealing the electron tunneling and the capture of the photo electrons. With a pulsed gate voltage and a pulsed laser intensity. In the shaded curves we observe the capture of optically generated electrons and the black solid line indicated the tunneling into the QD.}
           \label{fig4}
           \end{figure}

Finally, with another pulsed RF measurement, we are furthermore able to determine the tunneling rate into the QD without the influence of the photoinduced electron capture. We use a pulsed resonant excitation on the exciton transition in combination with a pulse gate voltage (Fig.~\ref{fig4}(a)) in a situation, where an electron will tunnel into the dot and quench the exciton transition. The gate voltage controls the tunneling while the pulsed laser switches on the photoinduced electron generation and probes simultaneously the RF signal from the exciton. In detail, we first prepare an empty QD state by setting the gate voltage to $V_1=0\,\text{V}$, a value well below the steady-state tunneling voltage at $0.45\,\text{V}$ (see also Fig.~\ref{fig1}(a)). The laser energy is adjusted so that RF will occur for a gate voltage $V_2=0.525\,\text{V}$, which lies above the tunneling voltage. At $t=0$, the gate voltage is switched to $V_2$ and shifts the QD exciton transition into resonance with the laser. A strong RF signal is visible at $t=0$ in Fig.~\ref{fig4}(b). The transition is now quenched by the two possible processes: (i) capture of photoinduced electrons and (ii) electron tunneling into the QD. The capture is faster than the tunneling, hence, for different time delays after the voltage pulse, we observe this process by fast exponential decay in the shaded area of the curves in Fig.~\ref{fig4}(b). At the beginning of each laser pulse, a strong RF intensity is visible as the electron capture has not set in yet. The maximum RF intensity for different time delays reflects the probability for tunneling, which happens on a slower time scale and visualized by the solid black line in Fig.~\ref{fig4}. We are able to determine the tunneling rate into the dot to $1\,\text{ms}^{-1}$ from this envelope function. This rate is a factor of two larger than the tunneling out process for the same gate voltage \cite{beckel2014asymmetry}

In conclusion, we have shown that even in resonant measurements (like RF and differential reflection) free, excited charge carriers are generated by excitation from a doped back contact. These charge carriers can be captured into the dot states and quench the exciton transition for a time duration that is given by the average tunneling time out of the dot. This effect is also present in samples with small tunneling barriers, where it is not observed in a quenching of the exciton transition, but leads to an increased linewidth, due the photoinduced electron in the vicinity of the dot and inside the dot. However, a possible influence on the linewidth needs further investigations in the future.

\end{document}